\documentstyle[12pt]{article}

\def\bc{\begin{center}}
\def\ec{\end{center}}
\def\be{\begin{equation}}
\def\ee{\end{equation}}
\def\noi{\noindent}

\def\del{\delta}
\def\al{\alpha}

\def\ffi{\varphi}

\def\kpp#1{{\bf k}'_{#1\perp}}
\def\b#1{{\bf #1}}
\def\ol#1{\overline{#1}}
\def\ra{\rightarrow}

\def\F#1{Fig.~\ref{f#1}}
\def\f#1{(\ref{#1})}
\def\lx1{x_1=\xi\rightarrow nx_0}
\def\x#1{x_#1\rightarrow 0}

\def\hs#1{\hspace{#1cm}}

\def\v#1#2#3#4{
\put(#1,#2){\circle*{2.00}}
\put(#1,#3){\circle*{2.00}}
\put(#1,#2){\dashbox{1.5}(0,#4)[t]{}}
}

\begin{document}

\bc
DEUTERON DISINTEGRATION IN QUARK-PARTON MODEL
\\
{\bf M.A.Braun and V.V.Vechernin}\\
{\it Department of Theoretical Physics,
St.Petersburg State University}\\
{\it 198904 St.Petersburg, Russia;
E-mail: vecherni@snoopy.phys.spbu.ru}
\ec

\begin{abstract}
The deuteron disintegration process with the emission of fast proton
in the vicinity of the kinematical boundary of the reaction,
when Feynman variable $x\rightarrow 2$, is studied.
The consideration is fulfilled in the framework of the
quark-parton model of cumulative phenomena based on
perturbative QCD calculations of the corresponding quark diagrams
near the thresholds, at which some quarks ("donors")
in the nuclear flucton transfer all their
longitudinal momenta to the distinguished active quarks and become soft.
The presence of the multi-quark $6q$-configuration in a deuteron
is essentially exploited in the consideration.
The different versions of hadronization mechanisms of the produced
cumulative quarks into cumulative particles are analyzed.
It is shown that in the case of the production of cumulative protons
from deuteron
the hadronization through the coalescence of three cumulative quarks
is favorable
and leads to the $(2-x)^5$ cross section threshold behavior whereas
the usual hadronization through one cumulative quark
fragmentation into proton
the same as the calculations predicts for
the deuteron structure function $F^d_2(x)$
at $x\rightarrow 2$ in DIS processes.
The results of the calculations are compared with the available
experimental data.
\end{abstract}

\section{Introduction}

The paper is devoted to the investigation of the
cumulative phenomena on deuteron at $x\rightarrow 2$
in terms of scaling variable $x$, defined for
the interaction with single nucleon.
This limit is of especial interest as
in the vicinity of the kinematical boundary of the reaction
the quantity $(2-x)$ can be considered as a small parameter
on which some perturbative scheme can be developed.
Note the uniqueness of deuteron in this respect,
it is practically impossible to hope to reach experimentally
the vicinity of the kinematical boundary $x\rightarrow A$
for any other nucleus.

It is clear that in this limit we have to use $6q$-language
for the description of deuteron, as in this case the typical internucleon
distances are smaller, than the mean nucleon radius.
The attempts to stay at the nucleon level of analysis lead to the
uncertainties in the so-called relativization procedure for the $NN$
wave function.
The origin of these difficulties is well known it is
the principal impossibility to develop the self-consistent theory
of the relativistic bound state with fixed number of interacting
constituents as in this case the particle number operator do not commutate
with the Hamiltonian.

The quark approach to the description of the processes on nuclei
is being developed during the long period of time (see, for example
\cite{Efr76}-\cite{Efr94}). In the paper \cite{Bro92}
it was pointed out on the possibility to use
the perturbative QCD calculations near the threshold,
at which quarks transfer all their
longitudinal momenta to the distinguished active quark and become soft.
In papers
\cite{NPB94}-\cite{JPG97}) we have applied
this idea for the description of the cumulative phenomena
at large $x$, $x\gg 1$.

\section{Obtained results}

In this paper we compare two mechanisms of the fast ($x\rightarrow 2$)
cumulative particle production in the deuteron disintegration
process.

The first mechanism is the production of one fast quark and
its subsequent fragmentation into cumulative particle.
This mechanism was studied in our papers \cite{NPB94}-\cite{JPG97},
as one responsible for the cumulative meson production.
For the process of deuteron fragmentation into pion
with $x_\pi\rightarrow 2$
it corresponds to the diagram of the type shown in \F 1.
Recall it's just the process from which the intensive
experimental investigations of cumulative phenomena
have been started \cite{Bal73}.
As $x_\pi\rightarrow 2$ then for
the longitudinal momentum of the fast quark we also have
$x\rightarrow 2$.  In result, all $x_i\rightarrow 0$ for
$i\geq 2$, which enable to apply the perturbative QCD scheme
of \cite{Bro92} for the calculation of this process.
As a result we find for the behavior of the inclusive
cross section
integrated over the transverse momentum
in the vicinity of the kinematical boundary of the reaction:
\be
I_{frag}^d(x) \sim (2-x)^9 \hs 1 {\rm at} \hs 1 x\rightarrow 2
\label{frag}
\ee
It's the same as for the behavior of the deuteron structure function
in the DIS process:
\be
F_2^d(x) \sim (2-x)^9 \hs 1  {\rm at} \hs 1  x\ra 2
\ee

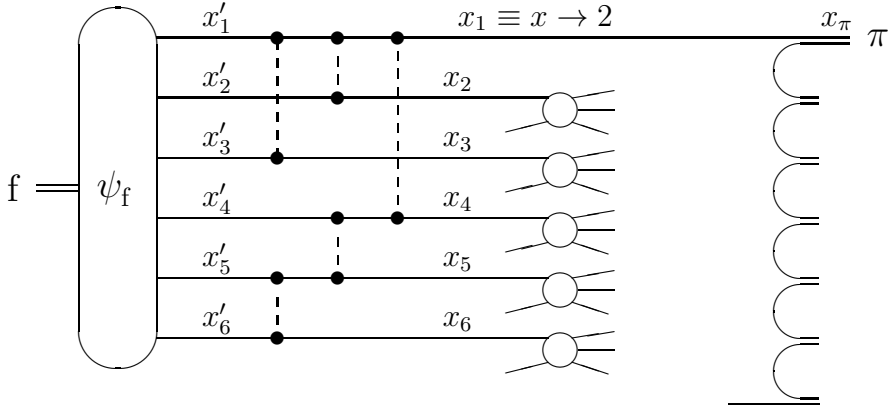
\begin{figure}

\unitlength=0.8mm
\linethickness{0.4pt}
\begin{picture}(150,70)(-20,0)

\multiput(30,10)(0,10){5}{\line(1,0){65}}

\put(30.00,60.00){\line(1,0){115.00}}
\put(10.00,34.50){\line(1,0){7.00}}
\put(10.00,35.50){\line(1,0){7.00}}

\put(7.00,35.00){\makebox(0,0)[rc]{\large f}}
\put(23.00,35.00){\makebox(0,0)[cc]{\large $\psi^{}_{\rm f}$}}
\put(148.00,60.00){\makebox(0,0)[lc]{\large $\pi$}}
\put(125.00,-1.00){\line(1,0){15.00}}
\put(140.00,59.00){\line(1,0){5.00}}
\put(40.00,62.00){\makebox(0,0)[cb]{$x'_1$}}
\put(40.00,52.00){\makebox(0,0)[cb]{$x'_2$}}
\put(40.00,42.00){\makebox(0,0)[cb]{$x'_3$}}
\put(40.00,32.00){\makebox(0,0)[cb]{$x'_4$}}
\put(40.00,22.00){\makebox(0,0)[cb]{$x'_5$}}
\put(40.00,12.00){\makebox(0,0)[cb]{$x'_6$}}
\put(143.00,62.00){\makebox(0,0)[cb]{$x_{\pi}$}}
\put(80.00,62.00){\makebox(0,0)[lb]{$x_1 \equiv x \ra 2$}}
\put(80.00,52.00){\makebox(0,0)[cb]{$x_2$}}
\put(80.00,42.00){\makebox(0,0)[cb]{$x_3$}}
\put(80.00,32.00){\makebox(0,0)[cb]{$x_4$}}
\put(80.00,22.00){\makebox(0,0)[cb]{$x_5$}}
\put(80.00,12.00){\makebox(0,0)[cb]{$x_6$}}
\v{50}{10}{20}{10}
\v{50}{40}{60}{20}
\v{60}{20}{30}{10}
\v{60}{50}{60}{10}
\v{70}{30}{60}{30}
\put(23.50,35.00){\oval(13.00,60.00)[]}

\multiput(0,0)(0,10){5}
{\begin{picture}(30,30)
\put(97.00,8.00){\circle{5.66}}
\put(95.00,6.00){\line(-4,-1){7.00}}
\put(99.00,6.00){\line(3,-1){6.00}}
\put(99.00,10.00){\line(6,1){7.00}}
\put(100.00,8.00){\line(1,0){6.00}}
\end{picture}}

\multiput(0,0)(0,10){6}
{\begin{picture}(30,30)
\put(140.00,4.50){\oval(15.00,9.00)[l]}
\end{picture}}

\end{picture}

\caption{
One fast quark production and
its fragmentation into cumulative pion.}
\label{f1}
\end{figure}

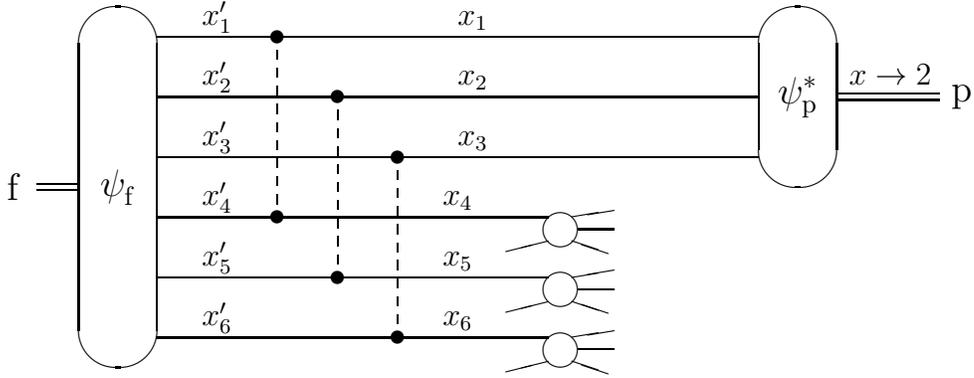
\begin{figure}

\unitlength=0.8mm
\linethickness{0.4pt}
\begin{picture}(150,70)(-20,0)

\multiput(30,10)(0,10){3}{\line(1,0){65}}
\multiput(30,40)(0,10){3}{\line(1,0){100}}

\put(143,49.5){\line(1,0){17}}
\put(143,50.5){\line(1,0){17}}
\put(10,34.5){\line(1,0){7}}
\put(10,35.5){\line(1,0){7}}

\put(7,35){\makebox(0,0)[rc]{\large f}}
\put(162,50){\makebox(0,0)[lc]{\large p}}
\put(40,62){\makebox(0,0)[cb]{$x'_1$}}
\put(40,52){\makebox(0,0)[cb]{$x'_2$}}
\put(40,42){\makebox(0,0)[cb]{$x'_3$}}
\put(40,32){\makebox(0,0)[cb]{$x'_4$}}
\put(40,22){\makebox(0,0)[cb]{$x'_5$}}
\put(40,12){\makebox(0,0)[cb]{$x'_6$}}
\put(145,52){\makebox(0,0)[lb]{$x \ra 2$}}
\put(80,62){\makebox(0,0)[lb]{$x_1$}}
\put(80,52){\makebox(0,0)[lb]{$x_2$}}
\put(80,42){\makebox(0,0)[lb]{$x_3$}}
\put(80,32){\makebox(0,0)[cb]{$x_4$}}
\put(80,22){\makebox(0,0)[cb]{$x_5$}}
\put(80,12){\makebox(0,0)[cb]{$x_6$}}
\v{50}{30}{60}{30}
\v{60}{20}{50}{30}
\v{70}{10}{40}{30}

\put(23.5,35){\oval(13,60)[]}
\put(23.5,35){\makebox(0,0)[cc]{\large $\psi^{}_{\rm f}$}}
\put(136.5,50){\oval(13,30)[]}
\put(136.5,50){\makebox(0,0)[cc]{\large $\psi^*_{\rm p}$}}

\multiput(0,0)(0,10){3}
{\begin{picture}(30,30)
\put(97,8){\circle{5.66}}
\put(95,6){\line(-4,-1){7}}
\put(99,6){\line(3,-1){6}}
\put(99,10){\line(6,1){7}}
\put(100,8){\line(1,0){6}}
\end{picture}}

\end{picture}

\caption{
Three fast quarks production and
their coalescence into cumulative proton.}
\label{f2}
\end{figure}

The second mechanism is the production of several fast quarks and
their coalescence into cumulative particle.
It was pointed out in our paper \cite{JPG97},
that in the case of the production of cumulative protons
the hadronization through the coalescence of three cumulative quarks
is favorable than the usual hadronization through one cumulative quark
fragmentation into proton.
For the process of deuteron fragmentation into proton
with $x\rightarrow 2$
it corresponds to the diagram in \F 2.
As $x=x_1+x_2+x_3 \ra 2$ then
all $x_i\rightarrow 0$ for
$i\geq 4$, which enable again to apply the perturbative QCD scheme
of \cite{Bro92} for the calculation of the process.
As a result we find for the behavior of the inclusive
cross section of the cumulative proton production
in the vicinity of the kinematical boundary of the reaction:
\be
I_{coal}^d(x) \sim (2-x)^5 \hs 1 {\rm at} \hs 1 x\rightarrow 2
\label{coal}
\ee

Preliminary comparison with the experimental data \cite{Abl83}
on the process of deuteron fragmentation into proton
in the vicinity of the kinematical boundary of the reaction,
at $x$ close to 2, shows that the behavior (\ref{coal}) of the
inclusive cross section corresponding to the coalescence mechanism
is compatible with the data and
the behavior (\ref{frag}) of the inclusive
cross section corresponding to the fragmentation mechanism
is incompatible with the data on cumulative proton production.

\section{Used approach and details}

First of all we shortly review the approximations made in the
process of diagram calculations.
In \F 1 $\psi_{\rm f}(x'_i,\kpp i)$ is the soft parton wave
function of the $6q$-flucton.
Following \cite{Bro92} we use hard gluon exchanges to calculate the
asymptotic behavior of the parton wave function at $x_1=x\ra 2 $
and $\x i,$ for $x_i\geq 2$.
Following \cite{Bro92} we choose the Coulomb gauge
in which transverse part of gluon exchanges is damped at low $x_i$
and the dominating Coulomb part is

\bc
\unitlength=0.5mm
\linethickness{0.4pt}
\begin{picture}(20,15)(70,5)
\multiput(10,10)(0,10){2}{\line(1,0){20}}
\put(7,10){\makebox(0,0)[rc]{$x'_2$}}
\put(7,20){\makebox(0,0)[rc]{$x'_1$}}
\put(33,10){\makebox(0,0)[lc]{$x_2$}}
\put(33,20){\makebox(0,0)[lc]{$x_1$}}
\put(50,15){\makebox(0,0)[cc]{=}}
\put(60,15){\makebox(0,0)[lc]{${\displaystyle
4\pi\al\frac{(x_1+x'_1)(x_2+x'_2)}{(x_1-x'_1)^2}\ .}$}}
\v{20}{10}{20}{10}
\end{picture}
\ec

For these light cone QCD calculations "old" time ordered perturbation
scheme is convenient. In this scheme we have for the "internal"
(between the gluon exchanges) quark propagators:

\bc
\unitlength=0.5mm
\linethickness{0.4pt}
\begin{picture}(60,15)(60,5)
\put(15,15){\line(1,0){20}}
\put(20,15){\circle*{2.00}}
\put(30,15){\circle*{2.00}}
\put(25,17){\makebox(0,0)[cb]{$x_i$}}
\put(45,15){\makebox(0,0)[cc]{=}}
\put(55,15){\makebox(0,0)[lc]{${\displaystyle \frac1{x_i}}$}}
\put(30,15){\dashbox{1.5}(0,10)[t]{}}
\put(20,5){\dashbox{1.5}(0,10)[t]{}}
\end{picture}
\ec

As a result in the framework of this scheme (or integrating over
$k'_{i-}$ and $k_{i-}$  momentum components in the usual Feynman
diagram approach) we find for the diagram in \F 1:

\be
I_{frag}^d(x) = \frac{4\pi^4}{9!}(2-x)^9 w_2 J_5
\left|\int
\ffi_{\rm f}(x'_1...x'_6)W_{frag}(x'_1...x'_6)
\del\left( \sum_1^6 x'_i-2 \right)
\prod_1^6 \frac{dx'_i}{2x'_i}
\right|^2
\label{frag1}
\ee
Here
\be
\ffi_{\rm f}(x'_i)\equiv\int\psi_{\rm f}(x'_i,\kpp i)
\del\left( \sum_1^6 \kpp i \right)
\prod_1^6 \frac{d\kpp i}{(2\pi)^3}
\label{fi}
\ee
The normalization condition is
\be
\int\left|\psi_{\rm f}(x'_i,\kpp i)\right|^2
\del\left( \sum_1^6 x'_i-2 \right)
\prod_1^6 \frac{dx'_i}{2x'_i}
\del\left( \sum_1^6 \kpp i \right)
\prod_1^6 \frac{d\kpp i}{(2\pi)^3}=1
\label{norm}
\ee
$W_{frag}$ includes gluon exchanges and
the "internal" quark propagators (see above).
$w_2$ is the probability to find the $6q$-state in deuteron.
$J_p$ describes the interactions of $p$ soft partons ("donors")
with the target.
\be
J_p=C^{-1}\int d^2 \b B \left[ 4\pi m^2 j(\b B)\right]^p
\label{Jp}
\ee
$C$ is the quasi-eikonal  factor, $m$ is the constituent quark mass and
the function j(\b B) has been calculated in \cite{NPB94} with eikonal
and in \cite{PAN97,JPG97} with quasi-eikonal
parametrization of the partonic amplitude.
It was demonstrated in \cite{NPB94} that all donors have to interact
with the target.

Following \cite{Bro92} we
assume that the soft partonic wave function
have the sharp maximum at
$x'_1=...=x'_6\equiv x'_0=2/6=1/3$. Then taking in \f{frag1} the function
$W_{frag}$ in this point we find the following approximation
for $I_{frag}^d(x)$:
\be
I_{frag}^d(x) = \frac{4\pi^4}{9!}(2-x)^9 w_2 J_5
\ol{W}^2_{frag}
\left|\int
\ffi_{\rm f}(x'_1...x'_6)
\del\left( \sum_1^6 x'_i-2 \right)
\prod_1^6 \frac{dx'_i}{2x'_i}
\right|^2
\label{frag2}
\ee
where
\be
\ol{W}_{frag}=
W_{frag}(x'_1=...=x'_6\equiv x'_0=1/3)=\frac{(4\pi\al)^5}{x'^4_0}X_5
\label{Wfrag}
\ee
and $X_5$ is the sum of about $10^2$ diagrams (in units of $x'_0$):

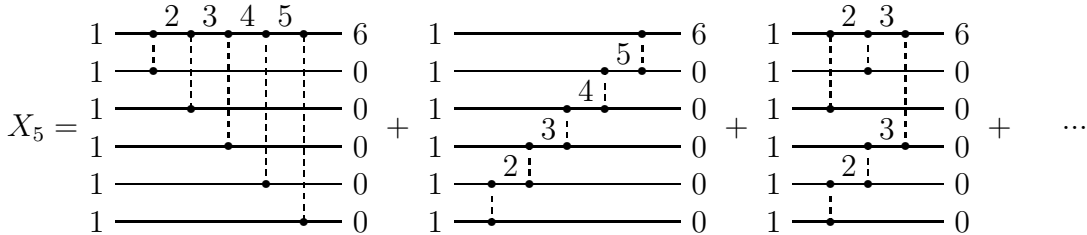
\unitlength=0.5mm
\linethickness{0.4pt}
\begin{picture}(200,70)(-20,5)

\put(0,0){
\begin{picture}(85,62)
\put(0,35){\makebox(0,0)[rc]{${\displaystyle X_5=}$}}
\multiput(10,10)(0,10){6}{\line(1,0){60}}
\multiput(7,10)(0,10){6}{\makebox(0,0)[rc]{$1$}}
\multiput(73,10)(0,10){5}{\makebox(0,0)[lc]{$0$}}
\put(73,60){\makebox(0,0)[lc]{$6$}}
\v{20}{50}{60}{10}
\v{30}{40}{60}{20}
\v{40}{30}{60}{30}
\v{50}{20}{60}{40}
\v{60}{10}{60}{50}
\put(25,62){\makebox(0,0)[cb]{$2$}}
\put(35,62){\makebox(0,0)[cb]{$3$}}
\put(45,62){\makebox(0,0)[cb]{$4$}}
\put(55,62){\makebox(0,0)[cb]{$5$}}
\put(85,35){\makebox(0,0)[cc]{+}}
\end{picture}
}
\put(90,0){
\begin{picture}(85,62)
\multiput(10,10)(0,10){6}{\line(1,0){60}}
\multiput(7,10)(0,10){6}{\makebox(0,0)[rc]{$1$}}
\multiput(73,10)(0,10){5}{\makebox(0,0)[lc]{$0$}}
\put(73,60){\makebox(0,0)[lc]{$6$}}
\v{20}{10}{20}{10}
\v{30}{20}{30}{10}
\v{40}{30}{40}{10}
\v{50}{40}{50}{10}
\v{60}{50}{60}{10}
\put(25,22){\makebox(0,0)[cb]{$2$}}
\put(35,32){\makebox(0,0)[cb]{$3$}}
\put(45,42){\makebox(0,0)[cb]{$4$}}
\put(55,52){\makebox(0,0)[cb]{$5$}}
\put(85,35){\makebox(0,0)[cc]{+}}
\end{picture}
}
\put(180,0){
\begin{picture}(85,62)
\multiput(10,10)(0,10){6}{\line(1,0){40}}
\multiput(7,10)(0,10){6}{\makebox(0,0)[rc]{$1$}}
\multiput(53,10)(0,10){5}{\makebox(0,0)[lc]{$0$}}
\put(53,60){\makebox(0,0)[lc]{$6$}}
\v{20}{10}{20}{10}
\v{30}{20}{30}{10}
\v{40}{30}{60}{30}
\v{20}{40}{60}{20}
\v{30}{50}{60}{10}
\put(25,22){\makebox(0,0)[cb]{$2$}}
\put(35,32){\makebox(0,0)[cb]{$3$}}
\put(25,62){\makebox(0,0)[cb]{$2$}}
\put(35,62){\makebox(0,0)[cb]{$3$}}
\put(65,35){\makebox(0,0)[cc]{+}}
\put(85,35){\makebox(0,0)[cc]{...}}
\end{picture}
}

\end{picture}

In \cite{NPB94} we have developed the method of these
diagram summation based on the recurrence relation for
the arbitrary number of quarks $X_p\equiv f_{p+1}(p+1)!$:
\be
f_{n}=\frac 1 {n(n-1)}\sum_{k=1}^{n-1}\frac{n+k}{n-k}f_{k}f_{n-k}
\label{fn}
\ee
with the initial condition $f_{1}=1$.
The recurrency relation (\ref{fn}) enables easy calculate $f_n$
for an arbitrary $n$ starting from $f_{1}=1$.
For large $n$ (\ref{fn}) evidently admits
asymptotical solutions of the form
\be
f_{n}\simeq [(6/5)n+o(n)]\exp (-an)
\label{fn1}
\ee
where $a$ is arbitrary. Numerical studies reveal that with $f_{1}=1$
\[ a=0.24421...\]
and also show that the asymptotical expression (\ref{fn1})
approximates the true
solution quite well starting from $n=3$, i.e. for all physically
interesting values.
In particular for $n=p+1=6$ we have
\be
X_5 = 6! f_6 = 6! \frac{36}{5}\exp (-1.464)
\label{X5}
\ee

The calculations of the diagram in \F 2 corresponding
to the mechanism of the coalescence of three fast quarks
into cumulative proton are very similar. As a result,
we find
$$
I_{coal}^d(x) = \frac{16\pi^2}{5!}(2-x)^5 w_2 J_3
\left|\int
\ffi_{\rm f}(x'_1...x'_6)W_{coal}(x'_1...x'_6,x_1,x_2,x_3)
\ffi^*_{\rm p}(x_1,x_2,x_3)
\right. \times
$$
\be
\times \left.
\del\left( \sum_1^6 x'_i-2 \right)
\prod_1^6 \frac{dx'_i}{2x'_i}
\del\left( \sum_1^3 x_i-2 \right)
\prod_1^3 \frac{dx_i}{2x_i}
\right|^2
\label{coal1}
\ee
We see that the interference takes place.
The result \f{coal1} is {\em not} reduced in general
to the product of two probabilities:
the probability to find three quarks with momenta $x_1,x_2,x_3$
multiplied by
the probability of the coalescence of quarks with these momenta.
We have to sum amplitudes not cross sections.
In \f{coal1} at first we have to integrate over $x_i$
and only then to calculate $|...|^2$.

The idea that in QCD a quark can hadronize by coalescing with
a comoving spectator parton was suggested in the paper
\cite{Gun87}.
It was used later for the
description of the fragmentation of protons and pions
into charm and beauty hadrons at large $x$
\cite{Vog92,Vog95}.
It was shown that the coalescence or recombination of
one or both intrinsic charm quarks with spectator valence
quarks of the Fock state leads in a natural way to
leading charm and beauty production.
But the interference effects were not taken
into account in these papers.

Again assuming that the soft partonic wave functions
have the sharp maximum at
$x'_1=...=x'_6\equiv x'_0=2/6=1/3$
and $x_1=x_2=x_3\equiv x_0=2/3$ we have:
$$
I_{coal}^d(x) = \frac{16\pi^2}{5!}(2-x)^5 w_2 J_3 \ol W^2_{coal} \times
$$
\be
\times
\left|\int
\ffi_{\rm f}(x'_1...x'_6)
\del\left( \sum_1^6 x'_i-2 \right)
\prod_1^6 \frac{dx'_i}{2x'_i}
\right|^2
\times
\left|\int
\ffi_{\rm p}(x_1,x_2,x_3)
\del\left( \sum_1^3 x_i-2 \right)
\prod_1^3 \frac{dx_i}{2x_i}
\right|^2
\label{coal2}
\ee
where
\be
\ol{W}_{coal}=
W_{coal}(x'_1=...=x'_6\equiv x'_0=1/3,x_1=x_2=x_3\equiv x_0=2/3)
=3!(4\pi\al X_1)
\label{Wcoal}
\ee
and $X_1$ is the simplest diagram (in units of $x'_0$):

\unitlength=0.5mm
\linethickness{0.4pt}
\begin{picture}(60,25)(-90,5)
\put(0,15){\makebox(0,0)[rc]{${\displaystyle X_1\ =}$}}
\put(40,15){\makebox(0,0)[lc]{${\displaystyle =\ 3}$}}
\multiput(10,10)(0,10){2}{\line(1,0){20}}
\multiput(7,10)(0,10){2}{\makebox(0,0)[rc]{$1$}}
\put(33,10){\makebox(0,0)[lc]{$0$}}
\put(33,20){\makebox(0,0)[lc]{$2$}}
\v{20}{10}{20}{10}
\end{picture}

\noi
3! corresponds to the time ordering of the gluon exchanges
in "old" perturbative scheme and no "internal" quarks
propagators enter in $W_{coal}$.

Note that related approach was successfully applied in \cite{Suk92}
for the description of the behavior of the nuclear structure
functions but in non cumulative $x<1$ region (the EMC effect).
In the paper the existence of a multi-quark cluster ($6q$)
was postulated and its structures functions were simply
approximated using the quark counting rules \cite{QCR73}
and normalization conditions.

We would like to emphasize that both
the intrinsic mechanism of the cumulative quark production
when the quarks of several nucleons concentrated
in one nuclear flucton transfer their
longitudinal momenta to the distinguished quark
and
the hadronization through the cumulative quarks coalescence
break the QCD factorization theorem.

This work is supported
by the Russian Foundation of Fundamental Research
under Grant No. 97-02-18123.

\end{document}